\documentclass[prb, twocolumn, longbibliography]{revtex4-2}
\usepackage{graphicx}
\usepackage{amsmath}
\usepackage{amssymb}
\usepackage{color}
\usepackage{appendix}
\usepackage{braket}
\usepackage{bbm}
\usepackage{subfigure}
\usepackage{mathrsfs}

\usepackage[colorlinks,urlcolor=blue,citecolor=blue,linkcolor=blue]{hyperref}
\usepackage{lipsum}
\usepackage{diagbox}
\usepackage{multirow}

\begin{document}
\title{Dynamical $\mathbb{Z}_{2}$ Skin Channels and Effective Loschmidt Cusps}
\author{Yongxu Fu}
\email{yongxufu@zjnu.edu.cn}
\affiliation{Department of Physics, Zhejiang Normal University, Jinhua 321004, China}

\begin{abstract}
We analytically describe the dynamically separated $\mathbb{Z}_{2}$ skin channels (wavepacket evolutions) under periodic boundary condition (PBC) in non-Hermitian systems with anomalous time-reversal symmetry (ATRS), by combining the semiclassical worldline perspective with an enhanced understanding of skin effects. These channels, tied to the initial state and relevant symmetries, exhibit individually exponential-dominated time evolution in momentum space, where their amplitude maxima evolve toward the dominant momenta. In real space, their center of masses (COMs) circulate around the one-dimensional (1D) chain, tracing semiclassical worldlines. Such circulations imply quantum revivals and effective Loschmidt cusps (LCs), with the latter showing scale-dependent behavior, a feature distinct from conventional dynamical quantum phase transitions. This work extends our previous findings on worldline windings and the winding-control mechanism to the $\mathbb{Z}_{2}$ skin effects, confirming that the core physics is shared with the ordinary skin effect.
\end{abstract}
\maketitle

\section{Introduction}
In non-Hermitian tight-binding lattices at the single-particle level, the ordinary skin effect (usually due to nonreciprocity) and the $\mathbb{Z}_{2}$ skin effect (e.g., arising from ATRS) are well-known hallmarks \cite{ashida2020, revmodphysNH, gong2018topo, kawabata2019symm, yao2018, yokomozi2019, zhang2020corres, okuma2020origin, kawabata2020symplectic}. The former is well described by non-Bloch band theory and the generalized Brillouin zone (GBZ) \cite{yao2018, yokomozi2019, zhang2020corres}, while the latter represents a symmetry-protected counterpart \cite{okuma2020origin, kawabata2020symplectic}. The ordinary skin effect has been extensively studied from a dynamical perspective \cite{longhi2022healing, xue2022burst, guo2022sticky, li2022dynamic, loghi2022self, zhu2024observation,  xiao2024burst, he2025anomalous}. In contrast, despite the recent detection of the $\mathbb{Z}_{2}$ skin effect in acoustic crystals \cite{wang2026acoustic}, its underlying physics and analytical characterization remain largely unexplored. This is especially true for the connection between the involved symmetries and the characteristic dynamical evolutions, as well as its subsequent manifestations.

In this paper, we combine the semiclassical worldline picture with an enhanced understanding of the two types of skin effects to provide an analytical description of the dynamically separated $\mathbb{Z}_{2}$ skin channels that appear under PBC of 1D non-Hermitian chain with ATRS. These channels are closely tied to the initial state (wavepacket compositions) and relevant symmetries. In momentum space, the amplitude maxima of the two $\mathbb{Z}_{2}$ skin channels each evolve toward their respective target momenta over time, while in real space, their COMs independently circulate around the chain, tracing out semiclassical worldlines. Furthermore, the circulating worldlines imply both amplitude revivals to the initial states and the emergence of effective LCs. Notably, such LCs exhibit scale-dependent behavior, a feature that distinguishes them from the conventional dynamical quantum phase transitions (DQPTs), which show genuine LCs in the thermodynamic limit \cite{heyl2013, review2018, zhou2018dqpts, modal2022dqpts, madal2023dqpts, zhang2025self, fu2025ana}. This work offers an intuitive demonstration of our previous quantum Monte Carlo stochastic series expansion (QMC-SSE) calculations of worldline winding numbers and the winding-control mechanism \cite{hu2023worldline, hu2025delocalization, fu2025windingcontrol}. Although the prior studies were conducted in the context of the ordinary skin effect, the core underlying physics is shared.

\section{Revisiting ordinary and $\mathbb{Z}_{2}$ skin effects}
Before proceeding, we revisit and deepen our understanding of the two types of skin effects in 1D non-Hermitian systems, which exhibit fundamental differences in two main aspects. First, at the level of bulk eigenstates under open boundary conditions (OBC), the ordinary skin effect allows a large number of bulk modes to localize at one end or at both ends of a 1D chain. This behavior is entirely determined by whether the GBZ intersects the unit circle, i.e., whether the modulus of $\beta$ over the entire GBZ is exclusively larger than $1$ (or smaller than $1$), or whether both cases coexist. In contrast, for the $\mathbb{Z}_{2}$ skin effect, due to ATRS, the OBC bulk eigenstates always appear as Kramers pairs, with each partner localizing at opposite ends of the chain. This feature can be understood from the symplectic extension of the non-Bloch band theory \cite{kawabata2020symplectic}. Second, from the perspective of topological origin, the ordinary skin effect is determined by whether the PBC spectrum exhibits a point gap, i.e., whether there exists a reference complex energy $\epsilon_{0}$ around which the PBC spectrum has a nonzero winding number. The spectral winding number is defined as
\begin{align}
    \label{eqwinding}
    W_{s}=\frac{1}{2 \pi i}\oint_{\mathrm{BZ}}\frac{d}{dk}\log \det \left[H(k)-\epsilon_{0}\right],
\end{align}
where $H(k)$ denotes the Bloch Hamiltonian. In contrast, for the $\mathbb{Z}_{2}$ skin effect, the spectral winding number is identically zero due to symmetry constraints. As a result, its topological characterization requires the introduction of a $\mathbb{Z}_{2}$ invariant \cite{okuma2020origin}.

In the ordinary skin effect, a subtle but often overlooked feature is that the OBC spectrum exists only in regions where the PBC spectral winding is nonzero. This feature becomes especially pronounced in more involved situations where the PBC spectral loops intersect, as we rigorously prove in the Supplemental Material \cite{supp}. For the $\mathbb{Z}_{2}$ skin effect, considering a two-band model as an example, the vanishing of $W_{s}$ in the presence of the $\mathbb{Z}_{2}$ skin effect arises because the two loop-type bands of $H(k)$, which form Kramers pairs, coincide with each other while running in opposite directions as $k$ varies. However, the nonzero winding number associated with a single band individually can indeed account for the skin effect. This can be achieved by employing the winding-control mechanism developed in our previous work to independently collapse a specific PBC spectral loop onto its corresponding OBC arcs (see Ref. \cite{fu2025windingcontrol} and the Supplemental Material \cite{supp} for details).

Building on the above conceptual insights, we now adopt a unified perspective on the two types of skin effects in terms of spectral winding numbers. Even though the total winding number vanishes for the $\mathbb{Z}_{2}$ skin effect, each individual band still features a winding number that plays a fundamental role in the underlying physics. According to the semiclassical picture, the winding number governs the unidirectional motion of quasiparticles, providing a dynamical physical picture of the skin effect that applies to arbitrary boundary conditions. Nevertheless, by combining the semiclassical understanding of the skin effect based on QMC-SSE worldlines \cite{hu2023worldline} with the winding-control mechanism \cite{fu2025windingcontrol}, we extend the analytical techniques developed for the wavepacket evolutions in the ordinary case \cite{he2025anomalous} and demonstrate the emergence of separated channels in the dynamical $\mathbb{Z}_{2}$ skin effect under symmetry constraints, as well as their consequential phenomena, e.g., scale-dependent LCs. This dynamical understanding is more simply realized in the ordinary skin effect, motivating our focus in the main text on the $\mathbb{Z}_{2}$ case, which has remained relatively unexplored.

\section{Dynamical $\mathbb{Z}_{2}$ skin channels}
We now consider a two-band non-Hermitian system belonging to the symplectic class, which exhibits ATRS such that
\begin{align}
    \mathcal{T}\mathcal{H}^{T}(k)\mathcal{T}^{-1}=\mathcal{H}(-k), \mathcal{T}\mathcal{T}^{*}=-1.
\end{align}
Under this symmetry, eigenstates at $k$ and $-k$ form Kramers pairs, implying for the two bands either $E_{\pm}(k)=E_{\pm}(-k)$ or $E_{\pm}(k)=E_{\mp}(-k)$ (see the Supplemental Material \cite{supp} for details). The former case implies that each band cannot form a loop in the complex plane, i.e., no point gap exists, and consequently no skin effect emerges, a scenario consistent with our winding-control mechanism \cite{fu2025windingcontrol}. The $\mathbb{Z}_{2}$ skin effect is compatible with the latter case, indicating that the two bands can coalesce into the same loop but traverse in opposite directions as $k$ varies. The loop carrying positive (negative) winding number indicates that the corresponding OBC eigenstates localized at the left (right) end of the chain, reflecting a negative (positive) imaginary velocity associated with a real Fermi surface $\mu$ that cuts vertically through the PBC loops. Such a point gap underlies the skin-driven dynamics in each band individually. Within the semiclassical picture, this imaginary velocity characterizes the average velocity of quasiparticles' worldlines, connecting to the factor $\exp[-i\mathcal{H}(k)t]$ that governs wavepacket dynamics through path integral or QMC-SSE calculations in imaginary time $\tau = it$ \footnote{This average velocity is understood as the time-averaged velocity over a sufficiently long interval, during which the dynamical driving underlying the skin effect is fully exhibited.}.

We proceed to investigate the dynamical behavior of a specific model. The symplectic Hatano-Nelson Hamiltonian preserving ATRS has the Bloch form (with the unit lattice constant)
\begin{align}
    \label{eq-sympHN}
    H_{s}(k)&=2t_h \cos k-2(\Delta \sigma_{x} +ig \sigma_{z})\sin k,
\end{align}
with $\mathcal{T}=\sigma_{y}$ \cite{okuma2020origin,kawabata2020symplectic}. The parameters $t_h$, $\Delta$, and $g$ are all real numbers. Besides, this model also obeys an additional pseudo-Hermiticity such that $\eta H_{s}^{\dagger}(k)\eta^{-1}=H_{s}(k)$, with $\eta=\sigma_{x}$. This necessitates the coexistence of complex conjugate eigenenergies, i.e., $E_{\pm}(k)=E_{\pm}^{*}(k)$ or $E_{\pm}(k)=E_{\mp}^{*}(k)$. The former regime ($|g|<|\Delta|$) preserves pseudo-Hermiticity, yielding an entirely real spectrum. The latter regime ($|g|>|\Delta|$) spontaneously breaks pseudo-Hermiticity, producing a complex spectrum in accordance with the point gap and $\mathbb{Z}_{2}$ skin effect. 
The $\mathbb{Z}_{2}$ topological invariant $\nu=0,1$ distinguishes the former and latter regimes, respectively (see the Supplemental Material \cite{supp} for detailed calculations), and is explicitly given by \cite{kawabata2019symm, okuma2020origin}
\begin{align}
    \label{eq-z2topoinvar}
    \left(-1\right)^{\nu}&=\mathrm{sgn}\Bigg\{ \frac{\mathrm{Pf}\left[\left(H_{s}(\pi)-E_{0}\right)\mathcal{T}\right]}{\mathrm{Pf}\left[\left(H_{s}(0)-E_{0}\right)\mathcal{T}\right]} \nonumber\\
    &\times \exp\left[-\frac{1}{2}\int_{0}^{\pi}\mathrm{d} \log \det \left[\left(H_{s}(k)-E_{0}\right)\mathcal{T}\right]\right] \Bigg\},
\end{align}
where $E_{0}$ denotes a reference point in the complex energy plane. Consequently, we focus on the latter scenario with $\nu=1$.

\begin{figure*}
    \centering
    \includegraphics[scale=1]{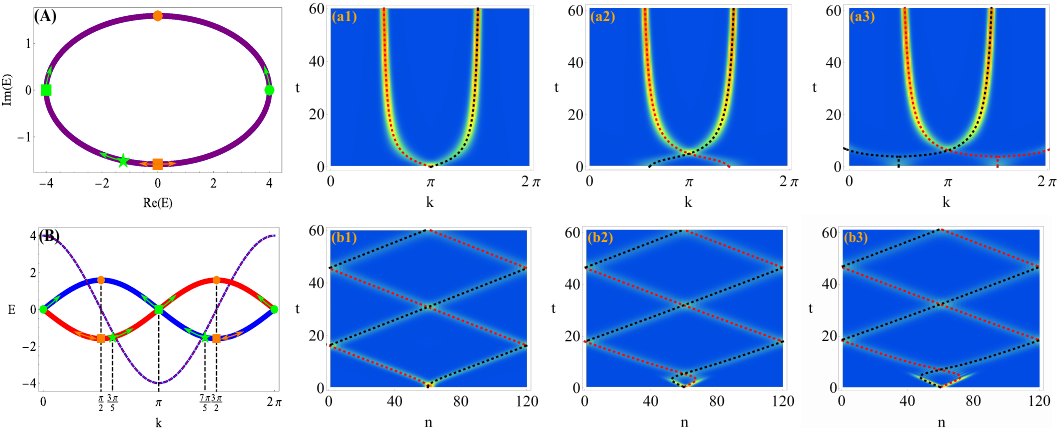}
    \caption{Dynamical $\mathbb{Z}_{2}$ skin channels of the symplectic Hatano-Nelson model in Eq. (\ref{eq-sympHN}). (A) PBC spectrum in the complex plane. (B) Real parts (blue and red dashed lines, overlapping) and imaginary parts (blue and red solid lines) of the two bands plotted against $k$. Green solid circles denote the Kramers degenerate point at $k=0$, while orange solid circles mark the maxima of the imaginary parts of the two bands. The three types of Gaussian wavepacket peak excitations are indicated by green solid squares, green stars, and orange solid squares, respectively. The green and orange arrows indicate the time-evolution directions of the respective initial momentum-peak positions. The corresponding evolutions of the dynamical $\mathbb{Z}_{2}$ skin channels, for the three types of excitations, are displayed in panels (a1)-(a3) for momentum space and in panels (b1)-(b3) for real space, respectively. The red and black dashed lines represent the corresponding analytical predictions from Eqs.~\eqref{eqkmaxz2} and~\eqref{eqcomz2} for the amplitude-maximum positions in the momentum-space panels (a1)-(a3) and the COM positions in the real-space panels (b1)-(b3), respectively. The blue background denotes the zero-amplitude region of the wavepacket. The wavepacket distributions in each panel are separately normalized at each time step. Without loss of generality, the relevant model parameters are taken as $t_h = 2$, $g = 0.8$, $\Delta = 0.1$, $\sigma = 0.4$, $\mathcal{C}_{\pm}=1/\sqrt{2}$, and $n_{0}=N/2$ with the number of lattice sites $N = 120$.}
    \label{fig-channelz2}
\end{figure*}

Given that the two PBC bands of this model merge into a single loop [depicted in purple in Fig. \ref{fig-channelz2}(A)], we construct an initial wavepacket of the form in the momentum space
\begin{align}
\label{eqinitial}
\ket{\psi(k,0)}=W_{k}^{+}\mathcal{C}_{+}\ket{u_{+}(k)}+W_{k}^{-}\mathcal{C}_{-}\ket{u_{-}(k)},
\end{align}
where $\ket{u_{\pm}(k)}$ are the right eigenstates of $H_{s}(k)$ associated with $E_{\pm}(k)$,
\begin{align}
\label{eqgaussian}
W_{k}^{\pm}=\exp \left[-\frac{(k-k_0^{\pm})^2}{2\sigma^{2}}-i(k-k_0^{\pm})n_{0}^{\pm}\right]
\end{align}
represent Gaussian wavepacket distributions of width $\sigma$, peaked at momentum $k_{0}^{\pm}$ and lattice site $n_{0}^{\pm}$ for the respective bands, and $\mathcal{C}_{\pm}\in [0,1]$ denote the participation coefficients \footnote{In this paper, we uniformly normalize the wavepacket evolutions in both momentum and real spaces at each instant. Hence, the initial Gaussian wavepacket components need not be explicitly normalized.}. 

For an initial excitation at generic $k_{0}^{\pm}$, the time-evolved wavepacket amplitude distribution \footnote{For notational brevity, throughout this paper, ``amplitude'' refers specifically to the squared modulus of the corresponding wave function.} of $\ket{\psi(k,t)} = \exp[-iH_{s}(k)t]\ket{\psi(k,0)}$ does not exhibit many intriguing features. By virtue of the ATRS of this model and the pseudo-Hermiticity breaking regime, characterized by $E_{\pm}(k)=E_{\mp}^{*}(k)$, the two bands evolve along separately wavepacket amplitude distributions, normalized at each time instant, which we refer to as two dynamical channels, provided that $k_{0}^{\pm}$ are set as a Kramers pair. This separation arises because the individual band components of the wavepacket amplitudes gradually dominate exponentially in time, while the cross terms between them are negligible due to broken pseudo-Hermiticity (see the Supplemental Material \cite{supp} for details). As the two separated channels are Kramers partners, we term them the dynamical $\mathbb{Z}_{2}$ skin channels, each driven in opposite directions with negative and positive (imaginary) average velocities, corresponding to the two counter-circulating PBC spectra with positive and negative winding numbers, respectively. By generalizing the methodology of Ref. \cite{he2025anomalous} to our $\mathbb{Z}_{2}$ skin effect scenario, the evolution of the $k$-space positions of the amplitude maxima $k_{max}^{\pm}$ for these two channels is governed by the self-consistent equations
\begin{align}
   \label{eqkmaxz2}
    k_{max}^{\pm}=k_{0}^{\pm}+\sigma^2 t \frac{d E_{\pm}^{I}(k)}{d k}\Bigg|_{k_{max}^{\pm}}.
\end{align}
The $k_{max}^{\pm}$ tend, as time evolves, to the points at which the imaginary part $E_{\pm}^{I}(k)$ of $E_{\pm}(k)=E_{\pm}^{R}(k)+iE_{\pm}^{I}(k)$ attains its maximum.
Upon Fourier transforming to real space, the two normalized dynamical $\mathbb{Z}_{2}$ skin channels remain separated, manifested as the separation of the COM positions $X_{c}^{\pm}(t)$ of their amplitude distributions, given by
\begin{align}
    \label{eqcomz2}
    X_{c}^{\pm}(t)=n_{0}^{\pm}+\mathcal{V}_{g}^{\pm}(t)t,
\end{align}
where
\begin{align}
    \mathcal{V}_{g}^{\pm}(t)&=\frac{\int dk\mathcal{A}_{\pm\pm}(k,k)\frac{dE_{\pm}^{R}(k)}{d k} e^{2E_{\pm}^{I}(k)t}}{\int dk \mathcal{A}_{\pm\pm}(k,k)e^{2E_{\pm}^{I}(k)t}}, \nonumber \\
    \mathcal{A}_{\alpha\beta}(k,k')&=W_{k}^{\alpha*}W_{k'}^{\beta}\mathcal{C}_{\alpha}^{*}\mathcal{C}_{\beta}\braket{u_{\alpha}(k)|u_{\beta}(k')}, \alpha,\beta=\pm. 
\end{align}
The total COM position of the two channels is equivalent, in terms of the exponential time evolution order, to a linear combination of $X_{c}^{\pm}(t)$ \cite{supp}, accompanied by a slow broadening of the wavepacket. The absence of exponentially growing time dependence in the interference terms between the two bands in the total COM position is a direct consequence of pseudo-Hermiticity breaking in the present model, thereby leading to the emergence of separated COMs of $\mathbb{Z}_{2}$ skin channels \cite{supp}. 

Generally, there are two conditions that ensure the emergence of two separated $\mathbb{Z}_2$ skin channels with proper initializations. Taking the present two-band model as an example, first, the two bands must each support skin-driven channels in both real and momentum spaces, with comparable exponential time scales. Second, the interference terms between them must be exponentially negligible in time compared with the independent channels. ATRS with nontrivial $\nu=1$ guarantees the first condition, as it ensures that the two Kramers-paired bands overlap with each other while winding in opposite directions, which allows us to design the initial wavepacket to satisfy the first condition. The spontaneous breaking of pseudo-Hermiticity, in turn, ensures the second condition by causing the exponentially suppressed interference channel. In contrast, for the trivial case $\nu=0$, there are no skin-driven wavepacket evolutions; the dynamics instead resemble the diffusive broadening typical of Hermitian systems. In this sense, the $\mathbb{Z}_2$ pair structure is robust and protected by ATRS, with the nontrivial invariant $\nu=1$ serving as the topological prerequisite for skin-driven dynamics in each band. The dynamical separation of the two channels, however, further requires appropriate initial-state preparation and suppression of inter-band interference. Hence, ATRS provides the topological condition, while the second condition governs the dynamical realization of two distinct $\mathbb{Z}_2$ skin channels.
In a broader context, for any two bands forming a Kramers pair (including but not limited to two-band models), one can always initialize the state according to Eq. (\ref{eqinitial}). Whenever the symmetry responsible for suppressing the exponentially growing time dependence in the interference terms is operative, e.g., pseudo-Hermiticity or parity-time symmetry, separated COMs of the $\mathbb{Z}_{2}$ skin channels are guaranteed to appear. Evidently, upon reducing from the $\mathbb{Z}_{2}$ skin effect scenario to the ordinary skin effect scenario, a single skin channel emerges, in agreement with both our semiclassical worldline picture and the winding-control mechanism \cite{supp}.

\section{Illustrations of dynamical $\mathbb{Z}_{2}$ skin channels and semiclassical worldlines}
Without loss of generality, we illustrate three types of $\mathbb{Z}_{2}$ skin channels corresponding to different initial momentum peaks $k_{0}^{\pm}$, with the same site peak $n_{0}$ for the model in Eq. (\ref{eq-sympHN}). The first case is specially taken at one of the two Kramers degenerate points $k_{0}^{\pm}=0,\pi$, namely $k_{0}^{\pm}=\pi$ (the other point behaves similarly), where $E_{+}(\pi)=E_{-}(\pi)$. This location is indicated by green solid squares in both the complex energy plane in Fig. \ref{fig-channelz2}(A) and the imaginary part of the dispersion (blue and red solid lines) in Fig. \ref{fig-channelz2}(B). The endpoints of the $k_{max}^{\pm}$ evolution are indicated by orange solid circles in Figs. \ref{fig-channelz2}(A) and \ref{fig-channelz2}(B). Notably, in Fig. \ref{fig-channelz2}(B), the imaginary parts of the two bands attain their maxima at distinct $k$ values, i.e., $\pi/2$ and $3\pi/2$, respectively \footnote{Throughout this work, we adopt the first BZ as $[0, 2\pi]$, under which the Kramers partner of a given $k$ is $2\pi - k$.}. Following Eq. (\ref{eqkmaxz2}), the energy points of the wavepacket peaks in the complex plane trace the fastest trajectory along the PBC spectral loop from their initial positions to the endpoint. This criterion is manifested in the dispersion relation, where the points on the two bands [blue and red bands in Fig. \ref{fig-channelz2}(B)] originate from the degenerate point at $\pi$ and move toward $\pi/2$ and $3\pi/2$ [green arrows in Fig. \ref{fig-channelz2}(B)], respectively. The momentum-space and real-space $\mathbb{Z}_{2}$ skin channels arising from such Kramers degenerate excitation are displayed in Figs. \ref{fig-channelz2}(a1) and \ref{fig-channelz2}(b1), with the initial real-space excitation peak $n_0$ located at the chain midpoint and the zero-amplitude distribution indicated by a blue background. The red and black dashed curves correspond to the analytical predictions for the two channels derived from Eqs. (\ref{eqkmaxz2}) and (\ref{eqcomz2}), respectively. As shown, after a short initial transient, these dashed lines precisely track the two channels. Here, owing to the PBC, the two real-space channels wrap around to the opposite end upon reaching the boundaries. Hence, in a semiclassical description, the two COM-evolution lines can be interpreted as a pair of worldlines encircling the system, offering an intuitive $\mathbb{Z}_{2}$ skin extension of our earlier QMC-SSE work on worldline winding \cite{hu2023worldline}.

In the second case, the wavepacket peak is excited at a generic Kramers pair [green star in Fig. \ref{fig-channelz2}(A)], corresponding to the two distinct momenta $k_{0}^{-}=3\pi/5$ and $k_{0}^{+}=7\pi/5$ for the two bands [Fig. \ref{fig-channelz2}(B)]. Nevertheless, as demonstrated in Figs. \ref{fig-channelz2}(a2) and \ref{fig-channelz2}(b2), the momentum-space and real-space $\mathbb{Z}_{2}$ skin channels remain well separated, and Eqs. (\ref{eqkmaxz2}) and (\ref{eqcomz2}) accurately capture the evolutions of the corresponding normalized wavepackets (red and black dashed lines). Importantly, the fastest-trajectory criterion governing the evolution of the dominant momentum from the initial $k_{0}^{\pm}$ to the endpoints remains valid, i.e., the wavepacket peaks travel along each band from the green stars to the orange solid circles [green arrows in Fig. \ref{fig-channelz2}(B)]. For our chosen initial momenta, the real-space $\mathbb{Z}_{2}$ channels undergo a directional reversal after a short initial evolution, in agreement with Eq. (\ref{eqcomz2}) aside from the details of the transient. Thereafter, the $\mathbb{Z}_{2}$ channels maintain their propagation direction and circulate along the chain under PBC, thereby tracing out two separated worldlines.  

In the third case, we consider a more subtle initial excitation, where the wavepacket peaks are located at $k_{0}^{-}=\pi/2$ and $k_{0}^{+}=3\pi/2$, the points at which the imaginary parts of the two bands attain their minima [orange solid squares in Figs. \ref{fig-channelz2}(A) and \ref{fig-channelz2}(B)]. Given the perfect symmetry of the two bands about their starting points, the fastest-trajectory criterion yields four momentum-space channels, visible shortly after the initial evolution in Fig. \ref{fig-channelz2}(a3). Although theory would require two of these channels to continue from the opposite end upon reaching the BZ boundary, numerical constraints from the finite BZ range and continuity prevent this from being resolved.  Nevertheless, this does not compromise the physical interpretation. Under these circumstances, the analytical expressions in Eq. (\ref{eqkmaxz2}) can be suitably adjusted to \footnote{Momentum-space channels in the full BZ follow from $2\pi$ translations of the first BZ.} accurately reproduce the numerical evolution paths [red and black dashed lines in Fig. \ref{fig-channelz2}(a3)]. In contrast, the real-space wavepacket evolution still yields two separated $\mathbb{Z}_{2}$ skin channels, which also correspond to semiclassical worldlines [Fig. \ref{fig-channelz2}(b3)], also in agreement with Eq. (\ref{eqcomz2}).

It should be clarified that the choice of real-space peaks $n_{0}^{\pm}$ or coefficients $\mathcal{C}_{\pm}$ in the initial wavepacket does not affect the emergence of $\mathbb{Z}_{2}$ skin channels. In particular, a deviation between $\mathcal{C}_{+}$ and  $\mathcal{C}_{-}$ only affects the relative amplitudes of the two channels. The only requirement for consistency between the two bands is the initial width $\sigma$; the excessive mismatch introduces a significant exponential difference in the amplitudes of the two channels. Notably, a larger Gaussian width $\sigma$ narrows the wavepacket in real space, thereby offering a faithful approximation to a single-site excitation \footnote{Large $\sigma$ merely approximates a single-site excitation; the clearest two-channel dynamics is achieved with suitably prepared wavepackets, whereas a generic localized excitation need not resolve two distinct propagating channels.}. Hence, tuning the above parameters readily yields a variety of experimentally accessible initial wavepackets. Meanwhile, spontaneous breaking of pseudo-Hermiticity is not the only way to achieve exponentially negligible inter-band interference; explicit breaking of pseudo-Hermiticity (e.g., via a global shift of the energy spectrum) can also render the interference terms exponentially negligible in time compared with the independent channels. Furthermore, even a weak explicit breaking of ATRS will exponentially destroy the separated $\mathbb{Z}_{2}$ skin channels. Further details of the above discussions are provided in the Supplemental Material \cite{supp}.

\begin{figure*}
    \centering
    \includegraphics[scale=1]{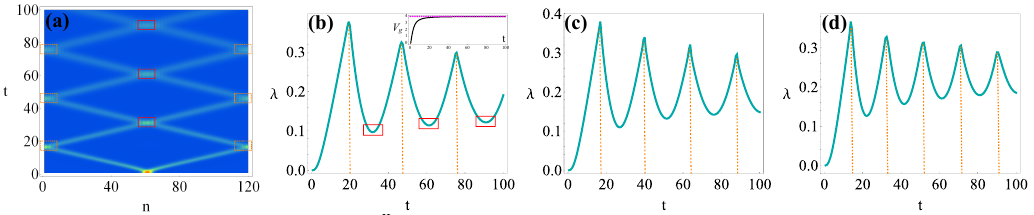}
    \caption{(a) Real-space $\mathbb{Z}_{2}$ skin channels, extended to longer times from Fig. \ref{fig-channelz2}(b1). The red and the orange open squares indicate the skin-driven quantum revivals and the effective LCs, respectively. (b) The corresponding rate function versus $t$ for $N=120$, where the orange dashed line marks the occurrence of effective LCs. The inset displays the time-asymptotic COM velocity $V_{g}$ of the two channels, which saturates at $4$ (magenta dashed line) under the parameters chosen in Fig. \ref{fig-channelz2}. (c)(d) Rate functions versus $t$ for $N=100$ and $N=80$, respectively, with other parameters same as in (b).}
    \label{fig-z2-dqpt}
    
\end{figure*}

\section{Skin-driven quantum revivals and effective LCs}
The $\mathbb{Z}_{2}$ skin channel phenomenon intuitively shows that the two channels and their COMs circulate along the 1D chain in real space as time evolves, returning nearly periodically to their in initial points, as indicated by the red open square in Fig. \ref{fig-z2-dqpt}(a), corresponding to an extension of Fig. \ref{fig-channelz2}(b1) to longer times. We refer to this behavior as quantum revival in the non-Hermitian sense. In a sufficiently long chain, the two channels propagate away from the initial point in opposite directions, causing their amplitude overlap with the initial wavepacket to steadily diminish, effectively approaching zero. However, as long as the chain remains finite, the channels inevitably return to their initial positions; consequently, the amplitude overlap first decreases and then grows again owing to the circulating motion at the boundaries.
This immediately suggests a hallmark of LCs in non-Hermitian scenario \cite{heyl2013, review2018, zhou2018dqpts, modal2022dqpts, madal2023dqpts, zhang2025self, fu2025ana}. To be specific, we introduce the finite-size Loschmidt echo and the rate function upon a self-norm formulation \footnote{For the initial state $\ket{\psi(0)}$ and the time-evolved state $\ket{\psi(t)}$, the self-normalization is implemented as $\ket{\psi(0)} / \sqrt{\braket{\psi(0)|\psi(0)}}$ and $\ket{\psi(t)} / \sqrt{\braket{\psi(t)|\psi(t)}}$, respectively.} 
\begin{align}
    \mathcal{L}(t) &= \left|\braket{\psi(0)|\psi(t)}\right|^{2},\\
    \lambda(t) &= -\frac{1}{N}\log \mathcal{L}(t),
\end{align}
where $\ket{\psi(t)}=\exp(-iHt)\ket{\psi(0)}$ denotes the wavepacket at time $t$, and $H$ (the model under consideration) is regarded as the Hamiltonian governing the quantum quench from the initial state $\ket{\psi(0)}$. We observe that in the vicinity of the time at which the wavepacket arrives at the boundary, namely, when $\mathcal{L}(t)$ reaches its minimum, signals an effective singularity in $\lambda(t)$, thereby identifying the occurrence of the effective LCs, as highlighted by the orange open squares in Fig. \ref{fig-z2-dqpt}(a) and the orange dashed lines in Fig. \ref{fig-z2-dqpt}(b) for the model in Eq. (\ref{eq-sympHN}). Here, an important observation is that, given the exponential disparity in wave function amplitudes at different locations, the influence of phase interference between $\bra{\psi(0)}$ and $\ket{\psi(t)}$ on $\mathcal{L}(t)$ becomes insignificant.

Nevertheless, such effective LCs considered here are finite-size revival-induced phenomena and should not be identified as conventional thermodynamic-limit DQPTs. The interval $\Delta t_{c}^{i}$ between two successive LCs $t_{c}^{i}$ and $t_{c}^{i+1}$ is estimated by
\begin{align}
    \left|\int_{t_{c}^{i}}^{t_{c}^{i+1}}\mathcal{V}_{g}^{\pm}(t)dt \right|= N.
\end{align}
Given the nearly linear wavepacket motion between successive LCs in Fig. \ref{fig-z2-dqpt}(a), the interval scales as $\Delta t_{c}^{i} \sim N / \bar{v}_{g}^{i}$. The magnitude of the average velocity $\bar{v}_{g}^{i}$ is comparable to the COM velocity $V_{g}=|\mathcal{V}_{g}^{\pm}(t)|$, which converges over time to the group velocity associated with the dominant momentum, as indicated by the magenta dashed line in the inset of Fig. \ref{fig-z2-dqpt}(b). Importantly, decreasing the system size reduces the interval between successive effective LCs, thereby increasing their number within the displayed time range [three, four, and five LCs, indicated by the orange dashed lines in Figs. \ref{fig-z2-dqpt}(b)-\ref{fig-z2-dqpt}(d), respectively]---a hallmark of scale-dependent behavior that distinguishes this scenario from thermodynamic-limit DQPTs. This scaling behavior applies to the early LCs; however, it gradually deviates as the wavepacket undergoes noticeable broadening over time.

From a semiclassical perspective, the physical origin of the quantum revivals and effective LCs discussed above can be traced to the cyclic motion of $\mathbb{Z}_{2}$ worldlines, separated worldlines endowed with nonzero winding numbers, akin to those in the ordinary skin effect \cite{hu2023worldline}. Accordingly, this skin-driven notion consistently applies to both the ordinary skin effect and its $\mathbb{Z}_{2}$ counterpart. This contrasts with the skin-free scenarios, particularly those where the COM undergoes Bloch oscillations \cite{he2025anomalous}; in such cases, the real-space wavepacket shows no skin-driven behavior and simply broadens over time. A similar analysis of skin-driven quantum revivals and effective LCs applies to the other parameter cases presented in Fig. \ref{fig-channelz2}, with the caveat that the wavepacket direction reversal during the early-time evolution must be taken into account.

\section{Discussions and Conclusions}
We have provided an analytical description of the dynamically separated $\mathbb{Z}_{2}$ skin channels under PBC in 1D non-Hermitian chains. The two channels are closely tied to the initial state and ATRS. The momentum-space amplitude maxima of the two channels evolve toward target momenta, while in real space their COMs circulate around the chain, forming semiclassical worldlines. These circulating worldlines give rise to finite-size revival-induced LCs, which exhibit scale-dependent behavior distinct from conventional DQPTs. This work also offers a demonstration of our previous QMC-SSE calculations on worldline circulations and the winding-control mechanism, confirming that the underlying physics is consistent with that of the ordinary skin effect. The $\mathbb{Z}_{2}$ skin channels are expected to be detectable in platforms such as acoustics crystals \cite{zhang2021acoustic, zhou2023observation, hu2025acoustic, wang2026acoustic}, photonic quantum walks \cite{xiao2020bbc, zhu2024observation, xiao2024burst, xue2024self}, and electric circuits \cite{lee2018topo, helbig2020, schindler2011exp, zhang2023electrical}. Under OBCs with disconnected ends on a 1D chain, the $\mathbb{Z}_{2}$ skin channels naturally travel to the boundaries without circulating around the chain. The detailed analysis of their boundary effects, such as reflection, may be informed by studies in the ordinary scenario \cite{li2022dynamic, he2025anomalous}. However, the difficulty lies in the fact that the $\mathbb{Z}_{2}$ case lacks a unified circular GBZ, rendering the global imaginary gauge transformation inapplicable. Resolving this issue represents a key challenge for future investigations, along with the further generalizations to higher dimensions and the incorporation of interactions.

\emph{Note added.---}
After completion of this work, we became aware of a recent related experimental work based on circuit simulation \cite{tang2026nonabelian}.

\section*{Acknowledgment}
Y.F. acknowledges support from the startup grant from Zhejiang Normal University (Grant No. YS304125915).

\newpage
\begin{widetext}

\begin{center}
\textbf{\large Supplemental Material for ``Dynamical $\mathbb{Z}_{2}$ Skin Channels and Effective Loschmidt Cusps"}
\end{center}

\section{Details of the eigenstates of the symplectic Hatano-Nelson model}
The symplectic Hatano-Nelson (HN) model studied in the main text is given by
\begin{align}
   \label{suppeq-syhn}
    H_{s}(k)=2t_h \cos k-2(\Delta \sigma_{x} +ig \sigma_{z})\sin k 
          =\left(\begin{matrix}
            2t_h \cos k-2ig \sin k & -2\Delta \sin k \\
            -2\Delta \sin k & 2t_h \cos k+2ig \sin k
        \end{matrix}\right),
\end{align}
which preserved ATRS $\sigma_{y}H^{T}_{s}(k)\sigma_{y}^{-1}=H_{s}(-k), \sigma_{y}\sigma_{y}^{*}=-1$.
The corresponding two band dispersions are expressed as
\begin{align}
    E_{\pm}(k)=2t_h \cos k \pm 2i\sqrt{g^2-\Delta^2}\sin k,
\end{align}
satisfying $E_{+}(-k)=E_{-}(k)$.
Solve the right eigenstates gives (up to normalization factors)
\begin{align}
  \label{suppeqstates}
  \ket{u_{\pm}(k)}\propto \left(\begin{matrix}
      i\Delta \\ g \pm \sqrt{g^2-\Delta^2}
  \end{matrix}\right),
\end{align}
which are independent to $k$ for such specific model. Nevertheless, our subsequent derivations are conducted using the general form $\ket{u_{\pm}(k)}$, so this particularity does not affect the general validity of our conclusions.
The left eigenstates $ \ket{\tilde{u}_{\pm}(k)}$ ($g>\Delta$) are givens by
\begin{align}
    H^{T}_{s}(k)&=H_{s}(k),\\
    H^{T}_{s}(k)\ket{\tilde{u}_{\pm}(k)}^{*}&=E_{\pm}(k)\ket{\tilde{u}_{\pm}(k)}^{*},\\
    \ket{\tilde{u}_{\pm}(k)}&=\ket{u_{\pm}(k)}^{*}\propto \left(\begin{matrix}
      -i\Delta \\ g \pm \sqrt{g^2-\Delta^2}
  \end{matrix}\right).
\end{align}
The biorthogonality normalization factors $\mathcal{N}_{\pm}$ are calculated by
\begin{align}
    \braket{\tilde{u}_{\pm}(k)|u_{\pm}(k)}=2(g^2-\Delta^2) \pm 2g \sqrt{g^2-\Delta^2},
\end{align}
although self-normalization can alternatively be employed as the situation demands.
Given that
\begin{align}
    E_{\pm}(k)\sigma_{y}\ket{\tilde{u}_{\pm}(k)}^{*}&=\sigma_{y}H^{T}_{s}(k)\ket{\tilde{u}_{\pm}(k)}^{*}=H_{s}(-k)\sigma_{y} \ket{\tilde{u}_{\pm}(k)}^{*},\\
    H_{s}(-k)\ket{u_{\pm}(-k)}&=E_{\pm}(-k)\ket{u_{\pm}(-k)}=E_{\mp}(k)\ket{u_{\pm}(-k)},\\
    E_{\pm}(-k)&=E_{\mp}(k),
\end{align}
which supports the point gap of PBC spectrum, thus,
\begin{align}
    H_{s}(-k)\ket{u_{\mp}(-k)}&=E_{\pm}(k)\ket{u_{\mp}(-k)} \propto E_{\pm}(k)\sigma_{y}\ket{\tilde{u}_{\pm}(k)}^{*},\\
    \ket{u_{\mp}(-k)}&\propto\sigma_{y}\ket{\tilde{u}_{\pm}(k)}^{*}=\sigma_{y}\ket{u_{\pm}(k)}.
\end{align}
Besides, the $\ket{u_{\mp}(-k)}$ are also given by
\begin{align}
    \ket{u_{\mp}(-k)}= \left(\begin{matrix}
      i\Delta \\ g \mp \sqrt{g^2-\Delta^2}
  \end{matrix}\right),
\end{align}
according to Eq. (\ref{suppeqstates}),
which should equivalent to 
\begin{align}
  \sigma_{y}\left(\begin{matrix}
      i\Delta \\ g \pm \sqrt{g^2-\Delta^2}
  \end{matrix}\right)=\left(\begin{matrix}
      0 & -i \\ i &0
  \end{matrix}\right)\left(\begin{matrix}
      i\Delta \\ g \pm \sqrt{g^2-\Delta^2}
  \end{matrix}\right)=\left(\begin{matrix}
      -i(g \pm \sqrt{g^2-\Delta^2}) \\ -\Delta
  \end{matrix}\right).
\end{align}
We carefully check the correctness of the corresponding relation as
\begin{align}
    \left(\begin{matrix}
      -i(g \pm \sqrt{g^2-\Delta^2}) \\ -\Delta
  \end{matrix}\right)=\frac{-(g \pm \sqrt{g^2-\Delta^2})}{\Delta}\left(\begin{matrix}
      i\Delta \\ \frac{\Delta^2}{(g \pm \sqrt{g^2-\Delta^2})}
  \end{matrix}\right)=\frac{-(g \pm \sqrt{g^2-\Delta^2})}{\Delta}\left(\begin{matrix}
      i\Delta \\ g \mp \sqrt{g^2-\Delta^2}
  \end{matrix}\right) \propto\ket{u_{\mp}(-k)}.
\end{align}
Hence, the Kramers pairs satisfy $E_{\pm}(k)=E_{\mp}(-k)=E_{\mp}^{*}(k)$, with the eigenstates related by $\sigma_{y}\ket{u_{\pm}(k)}\propto \ket{u_{\mp}(-k)}$. Kramers degeneracies occurs at $k=0$ and $k=\pi$.

\section{Uniform winding-control mechanism}
In this section, we demonstrate that the winding-control mechanism \cite{hu2023worldline, hu2025delocalization, fu2025windingcontrol} offers a unified account of both the ordinary and the $\mathbb{Z}_{2}$ skin effects. Around a reference complex energy $\epsilon_{0}$, the well-known periodic boundary condition (PBC) spectral winding number is defined as \cite{okuma2020origin, zhang2020corres}
\begin{align}
    \label{supp-eqwinding}
    W_{s}=\frac{1}{2 \pi i}\oint_{\mathrm{BZ}}\frac{d}{dk}\log \det \left[H(k)-\epsilon_{0}\right],
\end{align}
where $H(k)$ denotes the Bloch Hamiltonian. The existence of skin effects, particularly the ordinary one, is fundamentally related to nonzero winding numbers. The open boundary conditions (OBC) spectrum is confined to regions of nonzero PBC spectral winding, a subtle but often overlooked feature. This becomes especially pronounced in more complicated situations where the PBC spectral loops intersect. To substantiate such observation, we provide the following rigorous analysis. For the characteristic polynomial $f(\epsilon_{0},k)=\det \left[H(k)-\epsilon_{0}\right]$ given in Eq. (\ref{supp-eqwinding}), we perform an analytic continuation $e^{ik} \rightarrow z \in \mathbb{C}$ to obtain $f(\epsilon_{0},z)$, and assume that $z=0$ is a pole of order $p$. The integral in Eq. (\ref{supp-eqwinding}) is taken along the unit circle in the complex plane, and its result is given by the difference between the number of zeros and the number of poles (i.e., $p$) of $f(\epsilon_{0},z)$ enclosed by the contour. For $\epsilon_{0}$ lying in the region where $W_{s}$ is zero, among all zeros of $f(\epsilon_0, z)$ sorted in ascending order of their moduli, the $p$th and $(p+1)$th ones, denoted $z_p$ and $z_{p+1}$, lie inside and outside the unit circle, respectively. Consequently, the generalize Brillouin zone (GBZ) condition $|z_{p}|=|z_{p+1}|$ cannot be satisfied, and thus such $\epsilon_{0}$ cannot lie on the OBC spectrum. For $\epsilon_{0}$ in the region where $W_{s}$ is nonzero, $z_p$ and $z_{p+1}$ are either both inside or both outside the unit circle (corresponding to $W_{s}>0$ and $W_{s}<0$, respectively). Therefore, by continuously varying $\epsilon_{0}$ within this region, one can find points that satisfy the GBZ condition, namely the points on the OBC spectrum. 

For the $\mathbb{Z}_{2}$ skin effect, the anomalous time-reversal symmetry (ATRS) constrains the zeros of $f(\epsilon_{0},z)$ to appear in pairs $(z_{i}, z^{-1}_{i})$ located inside and outside the unit circle. We assume these zeros are ordered such that $|z_{1}|\leq \ldots \leq |z_{2lq}|\leq 1 \leq |z_{2lq}^{-1}|\ldots \leq|z_{1}^{-1}|$, where $l$ and $q$ denote the hopping range of the tight-binding model and the number of internal degrees of freedom per lattice site, respectively, and the number of poles of the characteristic polynomial is $2lq$. The symmetry constrains the total number of zeros (including accidental degeneracies) inside the unit circle to be $2lq$. These constraints force $W_{s}$ to be identically zero, rendering it ineffective as a topological characterization of the $\mathbb{Z}_{2}$ skin effect. In this case, the modified GBZ condition requires both $|z_{2lq-1}| = |z_{2lq}|$ and $|z_{2lq-1}^{-1}| = |z_{2lq}^{-1}|$, thus yielding a pair of GBZs \cite{kawabata2020symplectic}. Consequently, the corresponding bulk states become localized in pairs at both ends.

In our prior work \cite{fu2025windingcontrol}, the application of the winding-control mechanism to the ordinary skin effect reveals that PBC spectral loops with opposite winding numbers are subject to different boundary conditions. For a loop with a positive winding number, the imaginary velocity, derived from the Fermi sea occupation of the local PBC spectral loop under a Fermi surface that vertically traverses the spectrum, is negative, which in the semiclassical perspective corresponds to left-moving quasiparticles (worldlines). When the leftward hoppings at the chain ends are switched off, this loop collapses to the internal OBC counterpart, while rightward hoppings are irrelevant to this loop. An analogous treatment holds for loops with a negative winding number. This winding-control mechanism is valid for any complex model and for multiband situations. In the context of the $\mathbb{Z}_{2}$ skin effect, its applicability is treated independently for each of the two Kramers-degenerate loops, which we will illustrate below using the symplectic HN model in Eq. (\ref{suppeq-syhn}). The Brillouin zone (BZ) is known to correspond to the two oppositely winding PBC spectral loops of the two bands (purple circle and loop in Figs. \ref{supp-fig-cbcz2}(a) and \ref{supp-fig-cbcz2}(b), respectively), while the GBZ outer- and inner-unit circle components produce the degenerate two-band OBC spectrum (blue circle and line in the same figures). Further, by suppressing either the rightward or leftward hoppings at the chain ends under PBC, the corresponding clockwise (negative winding number) or counterclockwise (positive winding number) PBC spectral loop collapses to the respective OBC spectrum, while the other loop persists. Both cases give rise to the same green dotted spectral structure shown in Fig. \ref{supp-fig-cbcz2}(b). According to the physical meaning of the $\mathbb{Z}_{2}$ winding-control mechanism, the $\mathbb{Z}_{2}$ skin channels studied in the main text respectively correspond to the two PBC spectral loops of opposite winding directions. In real space, these correspond to left-moving and right-moving quasiparticles (worldlines or wavepackets).

\begin{figure}
    \centering
    \includegraphics[scale=1.2]{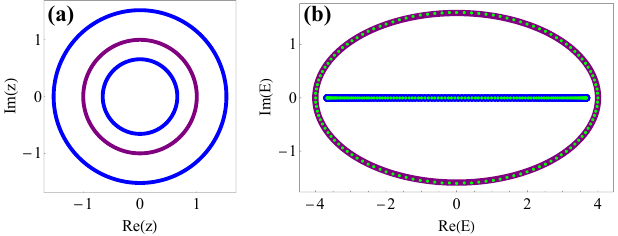}
    \caption{The application of the winding-control mechanism to the symplectic HN model in Eq. (\ref{suppeq-syhn}). Without loss of generality, the relevant model parameters are taken as $t_h = 2$, $g = 0.8$, $\Delta = 0.1$.}
    \label{supp-fig-cbcz2}
\end{figure}

\section{Explicit calculations of $\mathbb{Z}_{2}$ topological invariant}
Following the uniform winding-control mechanism discussed above, the condition for the $\mathbb{Z}_{2}$ skin effect is, intuitively, the point gap of the loop-type PBC spectrum. Even if the total winding number is zero, the bidirectionally localized skin modes arise from PBC bands with winding numbers of opposite signs. Alternatively, the $\mathbb{Z}_{2}$ skin effect can be  be quantified by the topological invariant given by \cite{kawabata2019symm, okuma2020origin}
\begin{align}
    \label{suppeq-z2topoinvar}
    \left(-1\right)^{\nu}=\mathrm{sgn}\left\{ \frac{\mathrm{Pf}\left[\left(H_{s}(\pi)-E_{0}\right)\mathcal{T}\right]}{\mathrm{Pf}\left[\left(H_{s}(0)-E_{0}\right)\mathcal{T}\right]} \times \exp\left[-\frac{1}{2}\int_{0}^{\pi}\mathrm{d} \log \det \left[\left(H_{s}(k)-E_{0}\right)\mathcal{T}\right]\right] \right\},
\end{align}
where $E_{0}$ denotes a reference point in the complex energy plane. With $\mathcal{T}=\sigma_{y}$ and $E_{0}=0$, we obtain 
\begin{align}
    \mathrm{sgn}\left\{ \frac{\mathrm{Pf}\left[\left(H_{s}(\pi)\right)\sigma_{y}\right]}{\mathrm{Pf}\left[\left(H_{s}(0)\right)\sigma_{y}\right]} \right\}=-1.
\end{align}
The other term in $(-1)^{\nu}$ can be evaluated via 
\begin{align}
    \label{suppexpcla}
    \exp\left[-\frac{1}{2}\int_{0}^{\pi}\mathrm{d} \log \det \left[H_{s}(k)\sigma_{y}\right]\right]=\sqrt{\frac{\det \left[H_{s}(0))\sigma_{y}\right]}{\det \left[H_{s}(\pi))\sigma_{y}\right]}},
\end{align}
which implies that $\nu$ is inherited from the $\mathbb{Z}_{2}$ topological invariant in the Hermitian case \cite{kawabata2019symm}.

On the other hand, the exponential factor can be explicitly evaluated by direct integration
\begin{align}
    \label{suppeq-inter}
    \int_{0}^{\pi}\mathrm{d} \log \det \left[H_{s}(k)\sigma_{y}\right]&=\int_{0}^{\pi}\mathrm{d} \log \det \left[S(k)\Lambda(k)S^{-1}(k)\sigma_{y}\right]\nonumber\\
    &=\int_{0}^{\pi}\mathrm{d} \log \det \left[S(k)\right]\det \left[\Lambda(k)\right]\det \left[S^{-1}(k)\right]\det \left[\sigma_{y}\right]
    \nonumber\\
    &=-\int_{0}^{\pi}\mathrm{d} \log \det \left[\Lambda(k)\right]\nonumber\\
    &=-\int_{0}^{\pi}\mathrm{d} \log (E_{+}(k)-0)(E_{-}(k)-0)\nonumber\\
    &=- \left[\int_{0}^{\pi}\mathrm{d}\log (E_{+}(k)-0)+\int_{0}^{\pi}\mathrm{d}\log(E_{-}(k)-0)\right],
\end{align}
where $\Lambda(k)=\left(\begin{matrix}
    E_{+}(k) & 0\\
    0 & E_{-}(k)
\end{matrix}\right)$ and $S(k)=\left(\begin{matrix}
    \ket{u_{+}(k)} & \ket{u_{-}(k)}
\end{matrix}\right)$.
For $|g|>|\Delta|$, in which the point gap exists, the sum of the half-winding-number integrals for the two bands in Eq. (\ref{suppeq-inter}) vanishes. As shown in Fig. \ref{supp-winding-z2}(a), this cancellation originates from the fact that the two bands share the same initial point at $k=0$, follow the upper and lower loop branches, respectively, and meet again at the same final point at $k=\pi$, resulting in half-winding numbers of opposite signs that cancel exactly. Consequently, $\nu=1$ in this nontrivial point-gap region. However, in the regime $|g|<|\Delta|$, the point gap collapses into an arc structure [Fig. \ref{supp-winding-z2}(b)], and the half-winding-number integral in Eq. (\ref{suppeq-inter}) becomes singular with respect to the reference point $E_{0}=0$. To regularize this singularity, we may shift the reference point slightly downward to $E_{0}=i0^{-}$. The integral in Eq. (\ref{suppeq-inter}) then yields $2\pi$, which corresponds to the trivial $\nu=0$. Therefore, the nontrivial $\mathbb{Z}_{2}$ point-gap regime studied in this work is protected by the topological invariant $\nu=1$.

\begin{figure}
    \centering
    \includegraphics[scale=0.5]{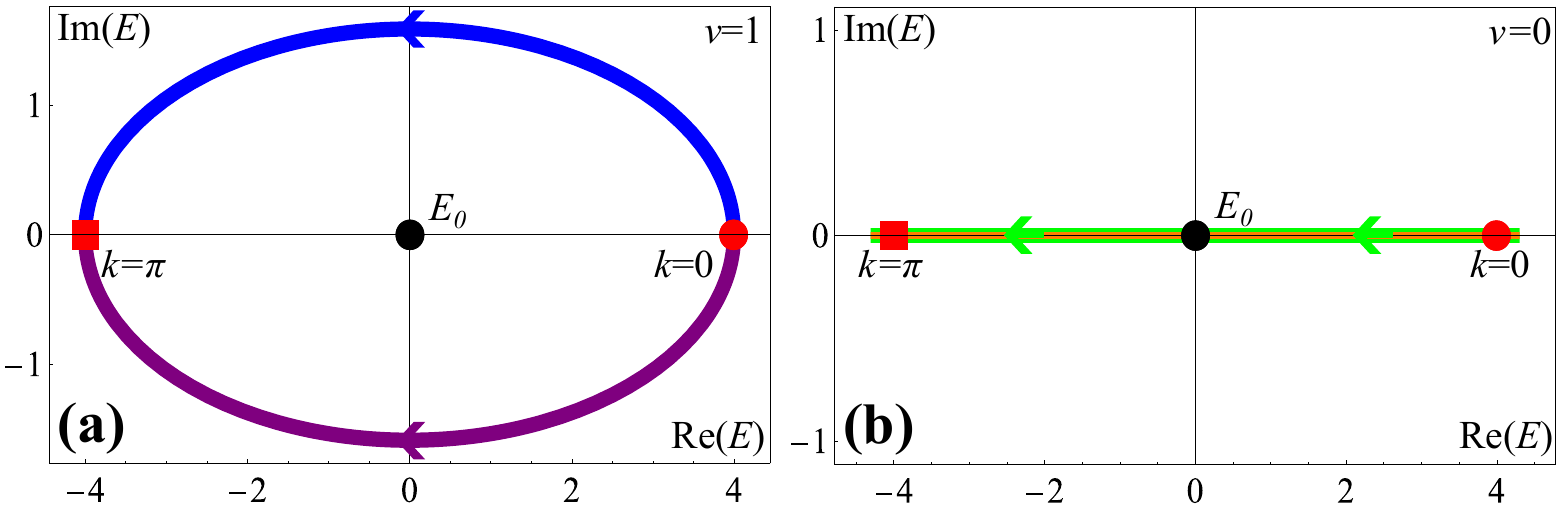}
    \caption{Explicit calculation of the $\mathbb{Z}_{2}$ topological invariant for the symplectic HN mode. (a) $\nu=1$ with $t_{h}=2, g=0.8, \Delta=0.1$, and $E_{0}=0$. The blue and purple curves denote $E_{+}(k)$ and $E_{-}(k)$, respectively. (b) $\nu=0$ with $t_{h}=2, g=0.1, \Delta=0.8$. The green and orange line segments represent $E_{+}(k)$ and $E_{-}(k)$, respectively. The reference point $E_{0}=i0^{-}$ is shifted downward by an infinitesimal amount relative to (a). Arrows on the bands indicate the direction of evolution from $k=0$ to $k=\pi$.}
    \label{supp-winding-z2}
\end{figure}

\section{Derivations of the dynamical $\mathbb{Z}_{2}$ skin channels}

\subsection{Momentum space }
According to the setup in the main text, we start from the initial Gaussian wavepacket amplitudes for the two bands of the pseudo-Hermiticity
broken symplectic system, with initial momentum peaks $k_{0}^{\pm}$ and real-space peaks $n_{0}^{\pm}$
\begin{align}
    W_k^{\pm}=\exp \left[-\frac{(k-k_0^{\pm})^2}{2\sigma^{2}}-i(k-k_0^{\pm})n_0^{\pm}\right],
\end{align}
respectively, which gives the initial state
\begin{align}
    \ket{\psi(k,0)}=W_{k}^{+}\mathcal{C}_{+}\ket{u_{+}(k)}+W_{k}^{-}\mathcal{C}_{-}\ket{u_{-}(k)}.
\end{align}
The corresponding time-evolution state is
\begin{align}
    \ket{\psi(k,t)}=e^{-i\mathcal{H}(k)t}\ket{\psi(k,0)}=W_{k}^{+}\mathcal{C}_{+}e^{-iE_{+}(k)t}\ket{u_{+}(k)}+W_{k}^{-}\mathcal{C}_{-}e^{-iE_{-}(k)t}\ket{u_{-}(k)}.
\end{align}
Denoting $E_{\pm}(k)=E_{\pm}^{R}(k)+iE_{\pm}^{I}(k)$, the amplitude evolution follows (up to an overall normalization coefficient over all $k$)
\begin{align}
    \left|\psi(k,t)\right|^2&=\left[W_{k}^{+*}\mathcal{C}_{+}^{*}e^{iE_{+}^{*}(k)t}\bra{u_{+}(k)}+W_{k}^{-*}\mathcal{C}_{-}^{*}e^{iE_{-}^{*}(k)t}\bra{u_{-}(k)}\right]\left[W_{k}^{+}\mathcal{C}_{+}e^{-iE_{+}(k)t}\ket{u_{+}(k)}+W_{k}^{-}\mathcal{C}_{-}e^{-iE_{-}(k)t}\ket{u_{-}(k)}\right]\nonumber\\
    &=\left|W_{k}^{+}\right|^2\left|\mathcal{C}_{+}\right|^2e^{2E_{+}^{I}(k)t}\braket{u_{+}(k)|u_{+}(k)}+\left|W_{k}^{-}\right|^2\left|\mathcal{C}_{-}\right|^2e^{2E_{-}^{I}(k)t}\braket{u_{-}(k)|u_{-}(k)}\nonumber\\
    &+W_{k}^{+*}W_{k}^{-}\mathcal{C}_{+}^{*}\mathcal{C}_{-}e^{i\left(E_{+}^{*}(k)-E_{-}(k)\right)t}\braket{u_{+}(k)|u_{-}(k)}+W_{k}^{-*}W_{k}^{+}\mathcal{C}_{-}^{*}\mathcal{C}_{+}e^{i\left(E_{-}^{*}(k)-E_{+}(k)\right)t}\braket{u_{-}(k)|u_{+}(k)}\nonumber\\
    &=\left|W_{k}^{+}\right|^2\left|\mathcal{C}_{+}\right|^2 e^{2E_{+}^{I}(k)t}\braket{u_{+}(k)|u_{+}(k)}+\left|W_{k}^{-}\right|^2\left|\mathcal{C}_{-}\right|^2 e^{2E_{-}^{I}(k)t}\braket{u_{-}(k)|u_{-}(k)} \nonumber\\
    &+W_{k}^{+*}W_{k}^{-}\mathcal{C}_{+}^{*}\mathcal{C}_{-}\braket{u_{+}(k)|u_{-}(k)}+W_{k}^{-*}W_{k}^{+}\mathcal{C}_{-}^{*}\mathcal{C}_{+}\braket{u_{-}(k)|u_{+}(k)},
\end{align}
where we have used the pseudo-Hermiticity breaking relation $E_{\pm}(k)=E_{\mp}^{*}(k)$.
Using the prescribed self-normalized right eigenstates $\braket{u_{\pm}(k)|u_{\pm}(k)}=1$, the expression simplifies to
\begin{align}
    \left|\psi(k,t)\right|^2=\left|W_{k}^{+}\right|^2\left|\mathcal{C}_{+}\right|^2 e^{2E_{+}^{I}(k)t}+\left|W_{k}^{-}\right|^2\left|\mathcal{C}_{-}\right|^2 e^{2E_{-}^{I}(k)t}+W_{k}^{+*}W_{k}^{-}\mathcal{C}_{+}^{*}\mathcal{C}_{-}\braket{u_{+}(k)|u_{-}(k)}+W_{k}^{-*}W_{k}^{+}\mathcal{C}_{-}^{*}\mathcal{C}_{+}\braket{u_{-}(k)|u_{+}(k)},
\end{align}
where the last two terms become negligible compared to the exponential dominant terms $e^{2E_{\pm}^{I}(k)t}$ as $t$ increases. Thus,
\begin{align}
    \left|\psi(k,t)\right|^2=\left|W_{k}^{+}\right|^2\left|\mathcal{C}_{+}\right|^2 e^{2E_{+}^{I}(k)t}+\left|W_{k}^{-}\right|^2\left|\mathcal{C}_{-}\right|^2 e^{2E_{-}^{I}(k)t}+\mathcal{O}(t^0).
\end{align}
We specify the two $\mathbb{Z}_{2}$ skin channels in the momentum space as
\begin{align}
    \mathcal{S}_{+} &= \left|\mathcal{C}_{+}\right|^2 e^{-(k-k_0^+)^2/\sigma^2+2E_{+}^{I}(k)t}
    =\left|\mathcal{C}_{+}\right|^2 e^{2X_{+}(t)},\\
    \mathcal{S}_{-} &= \left|\mathcal{C}_{-}\right|^2 e^{-(k-k_0^-)^2/\sigma^2+2E_{-}^{I}(k)t}
    =\left|\mathcal{C}_{-}\right|^2 e^{2X_{-}(t)}.
\end{align}
Following the derivations in Ref. \cite{he2025anomalous} and extending them to the $\mathbb{Z}_{2}$ scenario, the momentum-space amplitude peaks of the two channels are given by
\begin{align}
    0&=\frac{\partial X_{+}(t)}{\partial k}\Bigg|_{k_{max}^{+}}=\frac{k_0^+-k_{max}^{+}}{\sigma^2}+\frac{d E_{+}^{I}(k)}{d k}t\Bigg|_{k_{max}^{+}} \Rightarrow k_{max}^{+}=k_{0}^{+}+\sigma^2 t\frac{d E_{+}^{I}(k)}{d k}\Bigg|_{k_{max}^{+}},\\
    0&=\frac{\partial X_{-}(t)}{\partial k}\Bigg|_{k_{max}^{-}}=\frac{k_0^--k_{max}^{-}}{\sigma^2}+\frac{d E_{-}^{I}(k)}{d k}t\Bigg|_{k_{max}^{-}} \Rightarrow k_{max}^{-}=k_{0}^{-}+\sigma^2 t\frac{d E_{-}^{I}(k)}{d k}\Bigg|_{k_{max}^{-}}.
\end{align}

\subsection{Real space}
Using the Fourier transformation, the real-space wavepacket evolution is (up to an overall normalization coefficient over all $x$) 
\begin{align}
    \ket{\psi(n,t)}&=\frac{1}{\sqrt{2\pi}}\int dk \ket{\psi(k,t)}e^{ikn} \nonumber\\
    &=\frac{1}{\sqrt{2\pi}}\int dk \left[W_{k}^{+}\mathcal{C}_{+}e^{-iE_{+}(k)t}\ket{u_{+}(k)}+W_{k}^{-}\mathcal{C}_{-}e^{-iE_{-}(k)t}\ket{u_{-}(k)}\right]e^{ikn} \nonumber\\
    &=\frac{1}{\sqrt{2\pi}}\int dk \left[e^{-iE_{+}(k)t}\ket{\psi_{+}(k)}+e^{-iE_{-}(k)t}\ket{\psi_{-}(k)}\right]e^{ikn},
\end{align}
where 
\begin{align}
    \ket{\psi_{+}(k)}=W_{k}^{+}\mathcal{C}_{+}\ket{u_{+}(k)},\\
    \ket{\psi_{-}(k)}=W_{k}^{-}\mathcal{C}_{-}\ket{u_{-}(k)}.
\end{align}
and the amplitude is
\begin{align}
    \label{suppeq-realamp}
    &|\psi(n,t)|^{2}=\braket{\psi(n,t)|\psi(n,t)} \nonumber\\
                   &=\frac{1}{2\pi}\int dkdk' \left[e^{iE_{+}^{*}(k)t}\bra{\psi_{+}(k)}+e^{iE_{-}^{*}(k)t}\bra{\psi_{-}(k)}\right]e^{-ikn}\left[e^{-iE_{+}(k')t}\ket{\psi_{+}(k')}+e^{-iE_{-}(k')t}\ket{\psi_{-}(k')}\right]e^{ik'n} \nonumber\\
                   &=\frac{1}{2\pi}\int dk dk' e^{-i(k-k')n} \nonumber\\
                   &\times \left[\mathcal{A}_{++}(k,k')e^{i(E_{+}^{*}(k)-E_{+}(k'))t}+\mathcal{A}_{--}(k,k')e^{i(E_{-}^{*}(k)-E_{-}(k'))t}+\mathcal{A}_{+-}(k,k')e^{i(E_{+}^{*}(k)-E_{-}(k'))t}+\mathcal{A}_{-+}(k,k')e^{i(E_{-}^{*}(k)-E_{+}(k'))t}\right],
\end{align}
where
\begin{align}
    \mathcal{A}_{\alpha\beta}(k,k')&=\braket{\psi_{\alpha}(k)|\psi_{\beta}(k')}=W_{k}^{\alpha*}W_{k'}^{\beta}\mathcal{C}_{\alpha}^{*}\mathcal{C}_{\beta}\braket{u_{\alpha}(k)|u_{\beta}(k')}, \alpha,\beta=\pm, 
\end{align}
Our concern is the center of mass (COM) of the wavepacket
\begin{align}
    &X_{c}(t)=\frac{1}{\mathscr{N}(t)}\sum_{n}n\braket{\psi(n,t)|\psi(n,t)}=\frac{1}{\mathscr{N}(t)} \int dk dk' \frac{1}{2\pi} \sum_{n}ne^{-i(k-k')n} \nonumber\\
     &\times \left[\mathcal{A}_{++}(k,k')e^{i(E_{+}^{*}(k)-E_{+}(k'))t}+\mathcal{A}_{--}(k,k')e^{i(E_{-}^{*}(k)-E_{-}(k'))t}+\mathcal{A}_{+-}(k,k')e^{i(E_{+}^{*}(k)-E_{-}(k'))t}+\mathcal{A}_{-+}(k,k')e^{i(E_{-}^{*}(k)-E_{+}(k'))t}\right], \nonumber\\ 
\end{align}
where the time-evolved normalization coefficient is
\begin{align}
    &\mathscr{N}(t)=\sum_{n}\braket{\psi(n,t)|\psi(n,t)} =\int dk \left[\mathcal{A}_{++}(k,k)e^{2E_{+}^{I}(k)t}+\mathcal{A}_{--}(k,k)e^{2E_{-}^{I}(k)t}+\mathcal{A}_{+-}(k,k)+\mathcal{A}_{-+}(k,k)\right].
\end{align}
Using the formula
\begin{align}
    \frac{1}{2\pi} \sum_{n}ne^{-i(k-k')n}=i\frac{\partial\delta(k-k')}{\partial k},
\end{align}
the numerator of $X_{c}(t)$ is
\begin{align}
   &i \int dk dk' \frac{\partial\delta(k-k')}{\partial k} \nonumber\\
     &\times \left[\mathcal{A}_{++}(k,k')e^{i(E_{+}^{*}(k)-E_{+}(k'))t}+\mathcal{A}_{--}(k,k')e^{i(E_{-}^{*}(k)-E_{-}(k'))t}+\mathcal{A}_{+-}(k,k')e^{i(E_{+}^{*}(k)-E_{-}(k'))t}+\mathcal{A}_{-+}(k,k')e^{i(E_{-}^{*}(k)-E_{+}(k'))t}\right],
\end{align}
where the first term is
\begin{align}
    &i\int dk dk' \frac{\partial\delta(k-k')}{\partial k}\mathcal{A}_{++}(k,k')e^{i(E_{+}^{*}(k)-E_{+}(k'))t}=i\int dk \Bigg[\left(\delta(k-k')\mathcal{A}_{++}(k,k')e^{i(E_{+}^{*}(k)-E_{+}(k'))t}\right)\Bigg|_{k'-boundary}\nonumber\\
    &-\frac{\partial\mathcal{A}_{++}(k,k')}{\partial k}e^{i(E_{+}^{*}(k)-E_{+}(k'))t}\Bigg|_{k=k'}-i\mathcal{A}_{++}(k,k')\frac{d E_{+}^{*}(k)}{d k}t e^{i(E_{+}^{*}(k)-E_{+}(k'))t}\Bigg|_{k=k'}\Bigg]\nonumber\\
     &=n_{0}^{+}\int dk\left|W_{k}^{+}\right|^{2}\left|C_{+}\right|^{2}\braket{u_{+}(k)|u_{+}(k)}e^{2E_{+}^{I}(k)t}-i\int dk \left|W_{k}^{+}\right|^{2}\left|C_{+}\right|^{2}\braket{\frac{du_{+}(k)}{dk}|u_{+}(k)}e^{2E_{+}^{I}(k)t}\nonumber\\
     &+\int dk\left|W_{k}^{+}\right|^{2}\left|C_{+}\right|^{2}\braket{u_{+}(k)|u_{+}(k)}\frac{d E_{+}^{*}(k)}{d k}t e^{2E_{+}^{I}(k)t},
\end{align}
Using again the formula 
\begin{align}
    \frac{1}{2\pi} \sum_{n}ne^{-i(k-k')n}=-i\frac{\partial\delta(k'-k)}{\partial k'},
\end{align}
the numerator of $X_{c}(t)$ is also
\begin{align}
    &-i \int dk dk' \frac{\partial\delta(k'-k)}{\partial k'} \nonumber\\
     &\times \left[\mathcal{A}_{++}(k,k')e^{i(E_{+}^{*}(k)-E_{+}(k'))t}+\mathcal{A}_{--}(k,k')e^{i(E_{-}^{*}(k)-E_{-}(k'))t}+\mathcal{A}_{+-}(k,k')e^{i(E_{+}^{*}(k)-E_{-}(k'))t}+\mathcal{A}_{-+}(k,k')e^{i(E_{-}^{*}(k)-E_{+}(k'))t}\right],  
\end{align}
where the first term is also
\begin{align}
    &-i\int dk dk' \frac{\partial\delta(k'-k)}{\partial k'}\mathcal{A}_{++}(k,k')e^{i(E_{+}^{*}(k)-E_{+}(k'))t}=-i\int dk' \Bigg[\left(\delta(k-k')\mathcal{A}_{++}(k,k')e^{i(E_{+}^{*}(k)-E_{+}(k'))t}\right)\Bigg|_{k-boundary}\nonumber\\
    &-\frac{\partial\mathcal{A}_{++}(k,k')}{\partial k'}e^{i(E_{+}^{*}(k)-E_{+}(k'))t}\Bigg|_{k=k'}+i\mathcal{A}_{++}(k,k')\frac{d E_{+}(k')}{d k'}t e^{i(E_{+}^{*}(k)-E_{+}(k'))t}\Bigg|_{k=k'}\Bigg]\nonumber\\
     &=n_{0}^{+}\int dk\left|W_{k}^{+}\right|^{2}\left|C_{+}\right|^{2}\braket{u_{+}(k)|u_{+}(k)}e^{2E_{+}^{I}(k)t}+i\int dk \left|W_{k}^{+}\right|^{2}\left|C_{+}\right|^{2}\braket{u_{+}(k)|\frac{du_{+}(k)}{dk}}e^{2E_{+}^{I}(k)t}\nonumber\\
     &+\int dk\left|W_{k}^{+}\right|^{2}\left|C_{+}\right|^{2}\braket{u_{+}(k)|u_{+}(k)}\frac{d E_{+}(k)}{d k}t e^{2E_{+}^{I}(k)t}.
\end{align}
Since the eigenstates of the symplectic HN model of interest are independent of $k$, i.e., $\braket{\frac{du_{+}(k)}{dk}|u_{+}(k)}=\braket{u_{+}(k)|\frac{du_{+}(k)}{dk}}=0$, the first term of the numerator of $X_{c}(t)$ is
\begin{align}
    \frac{1}{2}\int dk\mathcal{A}_{++}(k,k)\left[2n_{0}^{+}+\frac{d (E_{+}(k)^{*}+E_{+}(k))}{d k}t\right] e^{2E_{+}^{I}(k)t}=\int dk\mathcal{A}_{++}(k,k)\left(n_{0}^{+}+\frac{dE_{+}^{R}(k)}{d k}t\right) e^{2E_{+}^{I}(k)t},
\end{align}
and similarly, the second term of the numerator of $X_{c}(t)$ is 
\begin{align}
   \int dk\mathcal{A}_{--}(k,k)\left(n_{0}^{-}+\frac{dE_{-}^{R}(k)}{d k}t\right) e^{2E_{-}^{I}(k)t}.
\end{align}
The third and fourth terms of the numerator of $X_{c}(t)$ can be calculated as
\begin{align}
     \mathrm{third \, term}=&i \int dk dk' \frac{\partial\delta(k-k')}{\partial k}\mathcal{A}_{+-}(k,k')e^{i(E_{+}^{*}(k)-E_{-}(k'))t}\nonumber\\
     =&i\int dk \left[\left(\delta(k-k')\mathcal{A}_{+-}(k,k')e^{i(E_{+}^{*}(k)-E_{-}(k'))t}\right)\Bigg|_{k'-boundary}-\frac{\partial\mathcal{A}_{+-}(k,k')}{\partial k}\Bigg|_{k=k'}
    -i\mathcal{A}_{+-}(k,k)\frac{d E_{+}^{*}(k)}{d k}t \right],\\
     \mathrm{fourth \, term}=&i \int dk dk' \frac{\partial\delta(k-k')}{\partial k}\mathcal{A}_{-+}(k,k')e^{i(E_{-}^{*}(k)-E_{+}(k'))t}\nonumber\\
     =&i\int dk \left[\left(\delta(k-k')\mathcal{A}_{-+}(k,k')e^{i(E_{-}^{*}(k)-E_{+}(k'))t}\right)\Bigg|_{k'-boundary}-\frac{\partial\mathcal{A}_{-+}(k,k')}{\partial k}\Bigg|_{k=k'}
    -i\mathcal{A}_{-+}(k,k)\frac{d E_{-}^{*}(k)}{d k}t \right].
\end{align}
Assuming the vanishing boundary terms, the dominant terms are the first and the second terms, thus
\begin{align}
    X_{c}(t)=\mathscr{N}_{+}(t)X_{c}^{+}(t)+\mathscr{N}_{-}(t)X_{c}^{-}(t)+\frac{2\mathrm{Re}\left[\mathcal{B}(t)\right]}{\mathscr{N}(t)},
\end{align}
where
\begin{align}
    \mathcal{B}(t)&=i\int dk \left(-\frac{\partial\mathcal{A}_{+-}(k,k')}{\partial k}\Bigg|_{k=k'}
    -i\mathcal{A}_{+-}(k,k)\frac{d E_{+}^{*}(k)}{d k}t\right), \\
    X_{c}^{+}(t)&=\frac{\int dk\mathcal{A}_{++}(k,k)\left(n_{0}^{+}+\frac{dE_{+}^{R}(k)}{d k}t\right) e^{2E_{+}^{I}(k)t}}{\int dk \mathcal{A}_{++}(k,k)e^{2E_{+}^{I}(k)t}}, \label{supeq-cen1} \\
    X_{c}^{-}(t)&=\frac{\int dk\mathcal{A}_{--}(k,k)\left(n_{0}^{-}+\frac{dE_{-}^{R}(k)}{d k}t\right) e^{2E_{-}^{I}(k)t}}{\int dk \mathcal{A}_{--}(k,k)e^{2E_{-}^{I}(k)t}}, \label{supeq-cen2} \\
\end{align}
are the separated COMs, and 
\begin{align}
    \mathscr{N}_{+}(t)&=\frac{\int dk \mathcal{A}_{++}(k,k)e^{2E_{+}^{I}(k)t}}{\mathscr{N}(t)}, \\
    \mathscr{N}_{-}(t)&=\frac{\int dk \mathcal{A}_{--}(k,k)e^{2E_{-}^{I}(k)t}}{\mathscr{N}(t)}.
\end{align}

We observe that while the real-space amplitude distributions in Eq. (\ref{suppeq-realamp}) show no explicit separation, i.e., the interference between the two bands is not negligible, the COM position are nevertheless dominated by the two $X_{c}^{\pm}(t)$ terms as time evolves. This behavior thus can be understood as the real-space manifestation of the $\mathbb{Z}_{2}$ skin channels, paralleling their momentum-space description.
Although the preceding derivation for this specific model benefits from the simplification that the eigenstates are independent of $k$, yielding a compact formula for the COM. The separated $\mathbb{Z}_{2}$ skin channels persist in generic cases, albeit with a more complex analytical expression for the COM, as long as the corresponding symmetry ensures that the cross terms between the channels (the third and fourth terms of $X_{c}(t)$) have no exponentially growing time dependence.

\section{Ordinary skin channels}
In this section, we present a typical ordinary skin channel example, with the Hamiltonian given by
\begin{align}
    \label{suppeq-hamord}
    H_{o}(k)=2\sin k-I\sin 2k.
\end{align}
As shown in Fig. \ref{supp-fig-ordinary}(a), the PBC spectrum features a self-intersection point (red point) and opposite winding numbers $\pm1$. We consider a specific initial Gaussian excitation to which we apply the theoretical $k$-peak $k_{max}$ and COM $X_{c}(t)$, the single-channel counterpart of our two-$\mathbb{Z}_{2}$-channel scenario. The momentum-space center $k_0=\pi/9$ is marked by the green star in Figs. \ref{supp-fig-ordinary}(a) and \ref{supp-fig-ordinary}(b), while the real-space center is set as $n_0=N/2$. The target momenta over time are the two maxima of $\mathrm{Im}[H_{o}(k)]$, indicated by the orange solid points in Figs. \ref{supp-fig-ordinary}(a) and \ref{supp-fig-ordinary}(b).

The presence of a self-intersection point gives rise to subtle features in the evolution of the momentum-space wavepacket when its peak approaches this point. In principle, the peak should follow the PBC spectral loop toward the target, following the green arrow leftward to the orange solid point in Fig. \ref{supp-fig-ordinary}(a), and along the $\mathrm{Im}[H_{o}(k)]$ curve from $\pi/9$ to $0$ and then from $2\pi$ to $7\pi/4$ in Fig. \ref{supp-fig-ordinary}(b). However, when the BZ range is taken as $[0,2\pi]$ in numerical simulations, the continuity requirement precludes a direct jump from $0$ to $2\pi$. Consequently, as the peak approaches $0$, the amplitude from $k=\pi$ progressively takes over, and the evolution effectively proceeds from $\pi$ to $3\pi/4$, as depicted by the rightward green arrow from the self-intersection point in Fig. \ref{supp-fig-ordinary}(a). This wavepacket evolution is illustrated in Fig. \ref{supp-fig-ordinary}(c). The black dashed line denotes the analytical peak position, while its mirror image about $0$ (red dashed line) roughly captures the peak evolution emanating from $\pi$. The real-space evolution after Fourier transformation is displayed in Fig. \ref{supp-fig-ordinary}(d), where the analytical COM result shows good agreement. Nevertheless, when the BZ zone is taken as $[-\pi, \pi]$, the wavepacket evolution indeed adheres to the anticipated trajectory, and the analytical results for the momentum-space peak as well as the real-space COM are in excellent agreement, as illustrated in Figs. \ref{supp-fig-ordinary}(e) and \ref{supp-fig-ordinary}(f). These two distinct wavepacket evolutions demonstrate that for spectral structures featuring self-intersection points, the appropriate BZ, rather than the fixed first BZ, is necessary. In the present case, the $[-\pi, \pi]$ BZ range provides the full BZ description. Furthermore, in the ordinary skin channel, the wavepacket undergoes relatively fast broadening as time evolves, in contrast to the $\mathbb{Z}_{2}$ skin channels, which broaden at a considerably slower rate.

\begin{figure}
    \centering
    \includegraphics[scale=1]{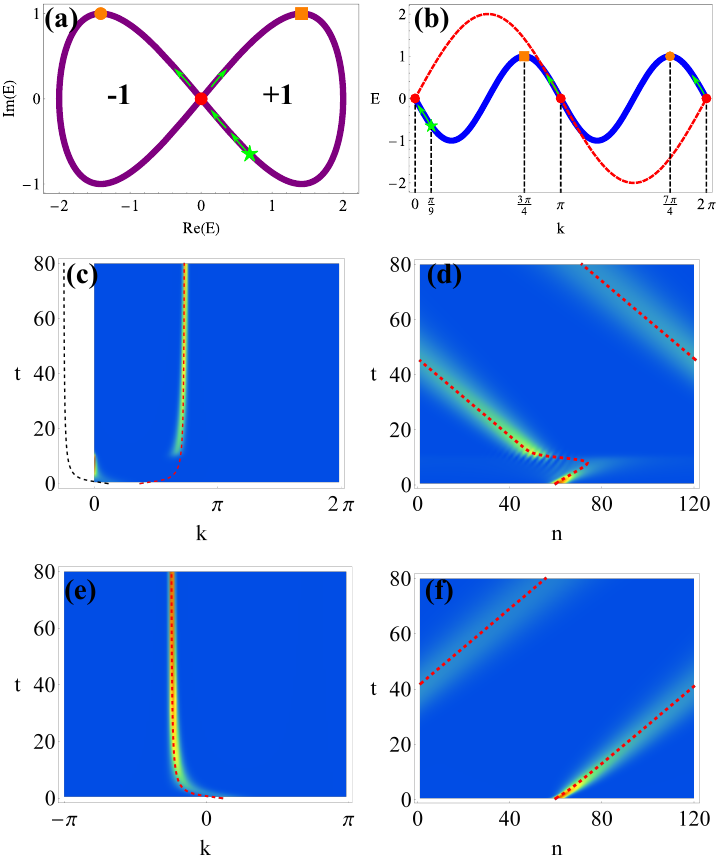}
    \caption{(a) PBC spectrum of the model in Eq. (\ref{suppeq-hamord}) with a self-intersection point (red dot) and opposite winding numbers $\pm1$. The green star marks the location of the initial momentum peak $k_0=\pi/9$ of the Gaussian wavepacket, and the orange solid points mark the maximum $\mathrm{Im}[H_{o}(k)]$. (b) Real (red dashed line) and imaginary (blue solid line) dispersion versus $k$, with the same markers. (c) Momentum-space evolution for BZ $[0,2\pi]$. Black dashed line is the analytical peak and red dashed line corresponds to its mirror about $0$, approximating the peak evolution from $\pi$. (d) Corresponding real-space evolution; analytical COM (black dashed) agrees well. (e)(f) Same as (c)(d) but for BZ $[-\pi,\pi]$, where the evolution follows the expected path. The width and initial real-space center of the Gaussian wavepacket are set to $0.4$ and $N/2$, respectively, with the number of lattice sites $N=120$.}
    \label{supp-fig-ordinary}
\end{figure}

\section{Further discussion of $\mathbb{Z}_{2}$ skin channels}
We further demonstrate the $\mathbb{Z}_{2}$ skin channels in more general scenarios, where the two channels have different initial centers $n_{0}^{\pm}$ of the Gaussian wavepackets in real space [Figs. \ref{supp-fig-n0dif}(a) and \ref{supp-fig-n0dif}(b)], and even different participation coefficients $\mathcal{C}_{\pm}$ for the two bands [Figs. \ref{supp-fig-n0dif}(c) and \ref{supp-fig-n0dif}(d)]. These results are also in full agreement with expectations. In particular, the two channels in real space are given by Eqs. (\ref{supeq-cen1}) and (\ref{supeq-cen2}) with different initial centers. The amplitude ratio of the two channels is fully determined by $\mathcal{C}_{\pm}$.

Furthermore, a width mismatch between the two components effectively introduces an exponential difference in the amplitudes of the two channels. For instance, by setting the two widths to $\sigma_{\pm} = 0.4 \pm 0.02$, i.e., deviating from each other by a small amount, the relative amplitude of the two channels already exhibits a dramatic difference [Fig. \ref{supp-fig-difwidth}]. One can envisage that a slightly larger deviation would cause one of the channels to effectively vanish. This deviation can be understood as follows. We unify the width of the two channels as $\sigma_-$, namely, approximately $e^{-(k - k_0^\pm)^2 / \sigma_-^2}$
Then the Gaussian wavepacket factor of the $+$ channel becomes $e^{-\frac{\sigma_-^2}{\sigma_+^2} \frac{(k - k_0^+)^2}{\sigma_-^2}}$, which is precisely the origin of the exponential difference. If the momentum-space width of the Gaussian wavepacket tends to infinity, the corresponding real-space wavepacket width vanishes, which is equivalent to a single-site excitation that is convenient for experimental implementation. In numerical simulations, this limit can be effectively approximated by taking sufficiently large values of $\sigma_{\pm}$. As shown in Fig.~\ref{supp-fig-larwidth}, taking $\sigma_{\pm} = 8$ effectively simulates the evolution of a single-site excitation, where the wavepacket is seen to spread over time.

\begin{figure}
    \centering
    \includegraphics[scale=1]{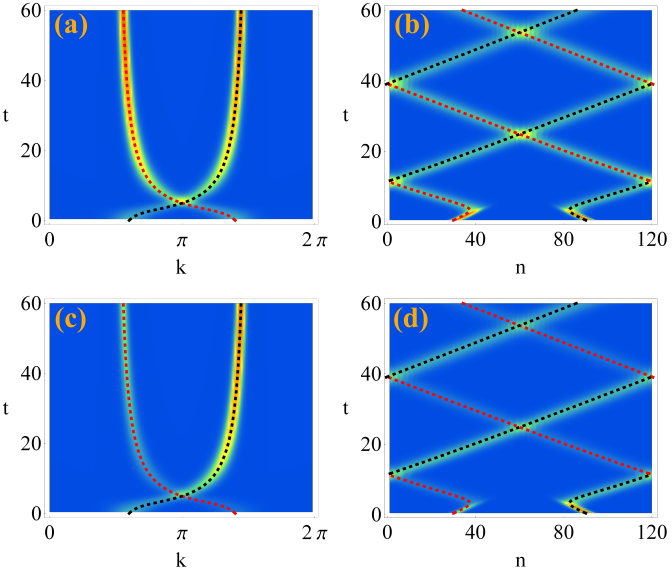}
    \caption{(a) and (b) show the $\mathbb{Z}_{2}$ skin channels for the model presented in the main text [or Eq. (\ref{suppeq-syhn})], with different initial real-space centers for the two-band components of the Gaussian wavepackets: $n_{0}^{+}=N/4$, $n_{0}^{-}=3N/4$. (c) and (d) further show the case where the participation coefficients are also different: $\mathcal{C}_{+}=0.5/\sqrt{0.5^2+0.8^2}$, $\mathcal{C}_{-}=0.8/\sqrt{0.5^2+0.8^2}$. The analytical results (red and black dashed lines) accurately track the channels in both momentum and real spaces. Other parameters are the same as those in the main text.}
    \label{supp-fig-n0dif}
\end{figure}

\begin{figure}
    \centering
    \includegraphics[scale=0.22]{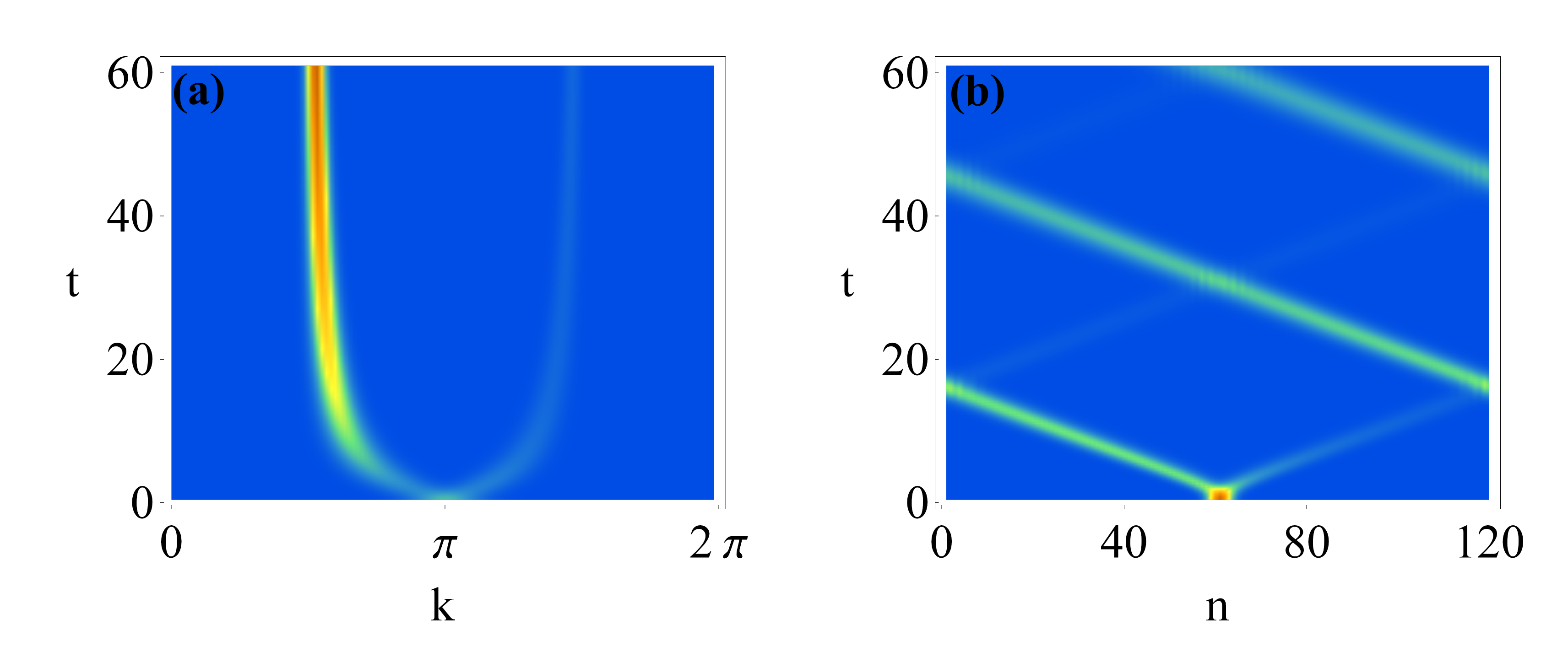}
    \caption{(a) and (b) show the $\mathbb{Z}_{2}$ skin channels for the model presented in the main text [or Eq. (\ref{suppeq-syhn})], with different widths $\sigma_{\pm}=0.4 \pm 0.02$ for the two-band components of the Gaussian wavepackets. Other parameters are the same as those in the main text.}
    \label{supp-fig-difwidth}
\end{figure}

\begin{figure}
    \centering
    \includegraphics[scale=0.22]{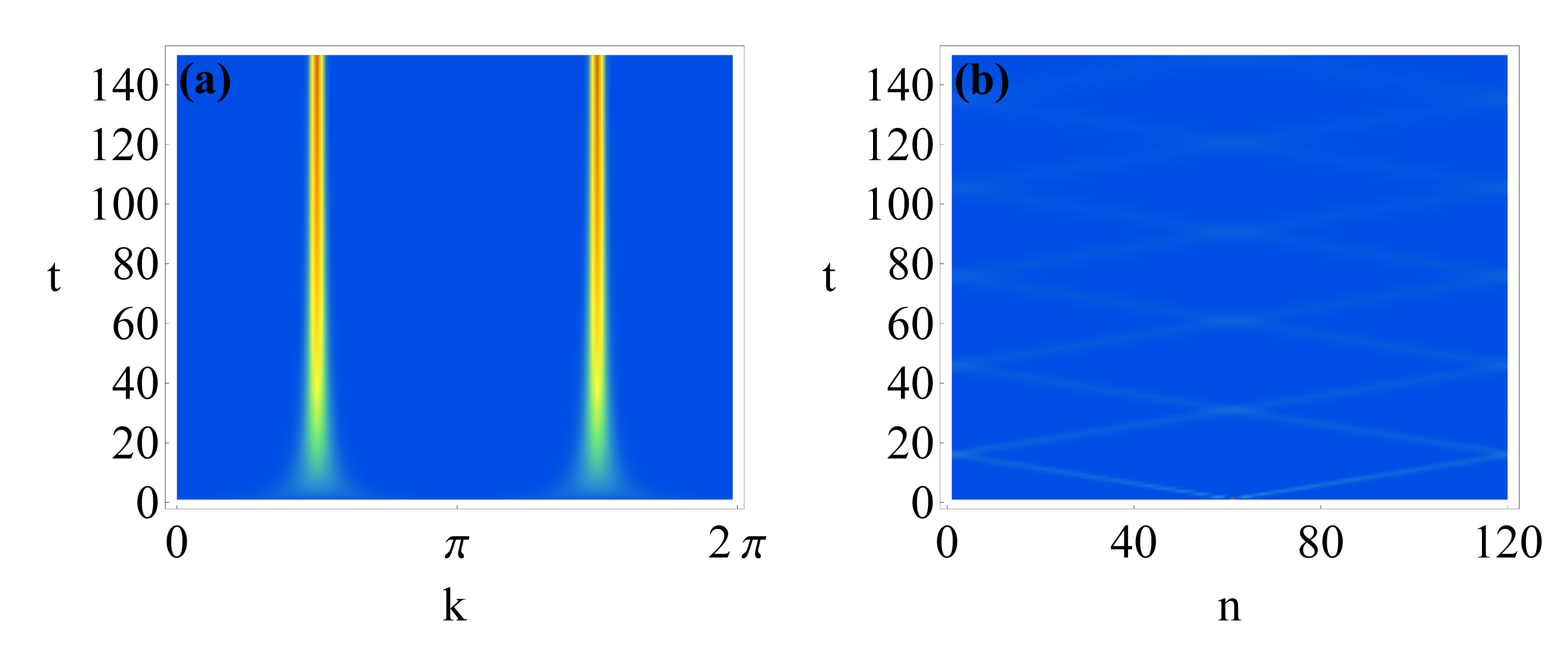}
    \caption{(a) and (b) show the $\mathbb{Z}_{2}$ skin channels for the model presented in the main text [or Eq. (\ref{suppeq-syhn})], with larger momentum-space width $\sigma_{\pm}=8$ for the two-band components of the Gaussian wavepackets. Other parameters are the same as those in the main text.}
    \label{supp-fig-larwidth}
\end{figure}

\section{Explicit breaking of ATRS}
ATRS is a prerequisite for obtaining dynamical $\mathbb{Z}_2$ skin channels by properly setting the initial excitation. In other words, if ATRS is explicitly broken, the original $\mathbb{Z}_2$ skin-channel structure is also destroyed. To demonstrate explicitly, we break ATRS in Eq. (\ref{suppeq-syhn}) by adding $\lambda \sigma_{x}$, giving
\begin{align}
    \label{supeq-sybre}
    H_b(k) = H_s(k) + \lambda \sigma_x.
\end{align}
The resulting split Kramers-pair spectra and wavepacket dynamical evolution are shown in Fig. \ref{supp-fig-symbre} with $\lambda=0.2$. The magnitude of $\lambda$ controls the degree of splitting of the two coincident energy spectra under ATRS. As a result, in the wavepacket dynamics, the exponential dominance of the skin channel corresponding to one of the bands becomes progressively enhanced over time, eventually leaving a single ordinary skin channel. It is noteworthy that even a weak breaking of ATRS can lead to an individual enhancement that is not negligible over time. Moreover, when $\lambda$ becomes sufficiently large (roughly greater than $1.45$), both the steady-state skin effect and the dynamical skin channels cease to manifest, and the wavepacket spreads out in real space.

\begin{figure}
    \centering
    \includegraphics[scale=0.21]{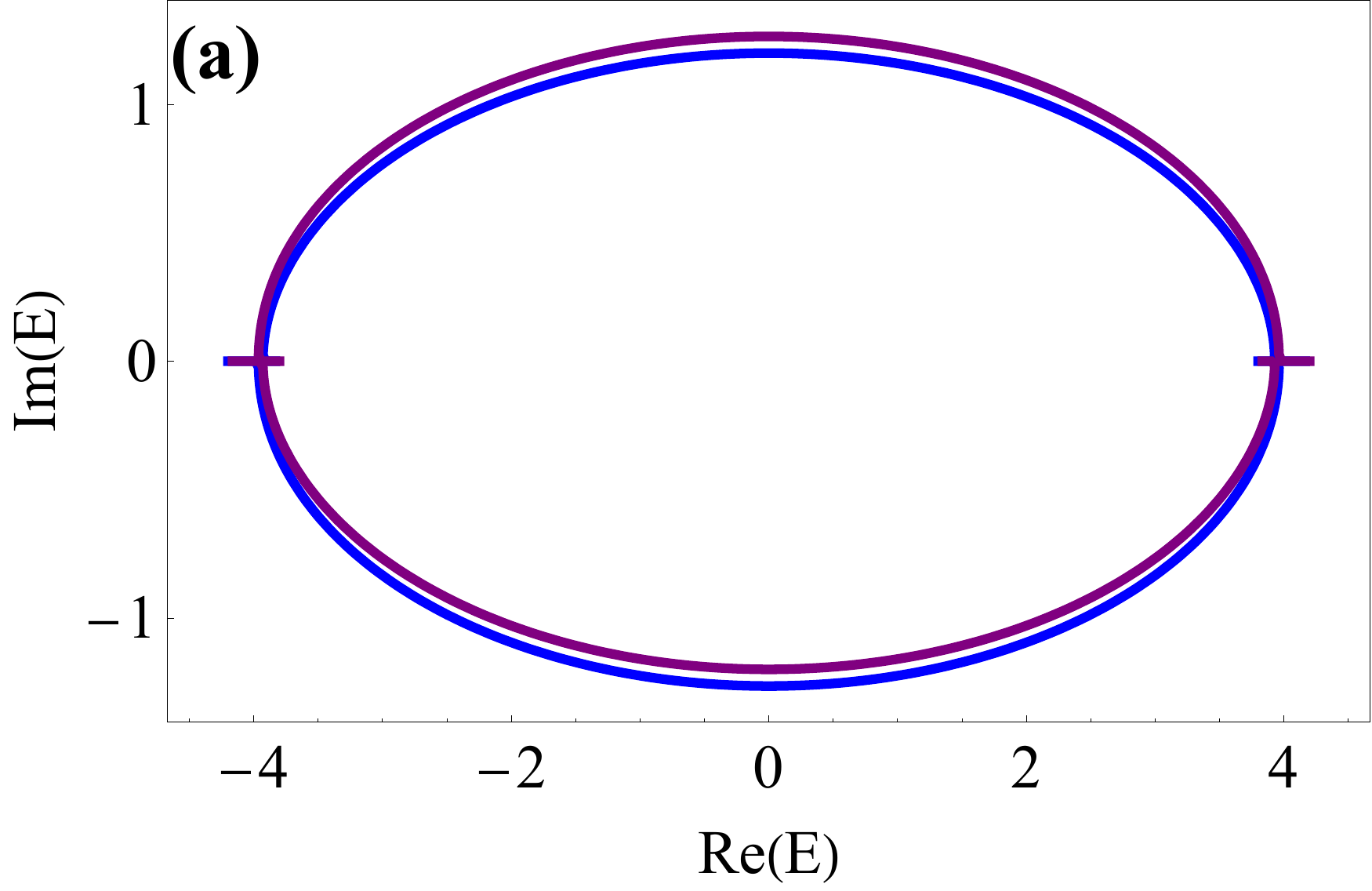}
    \includegraphics[scale=0.195]{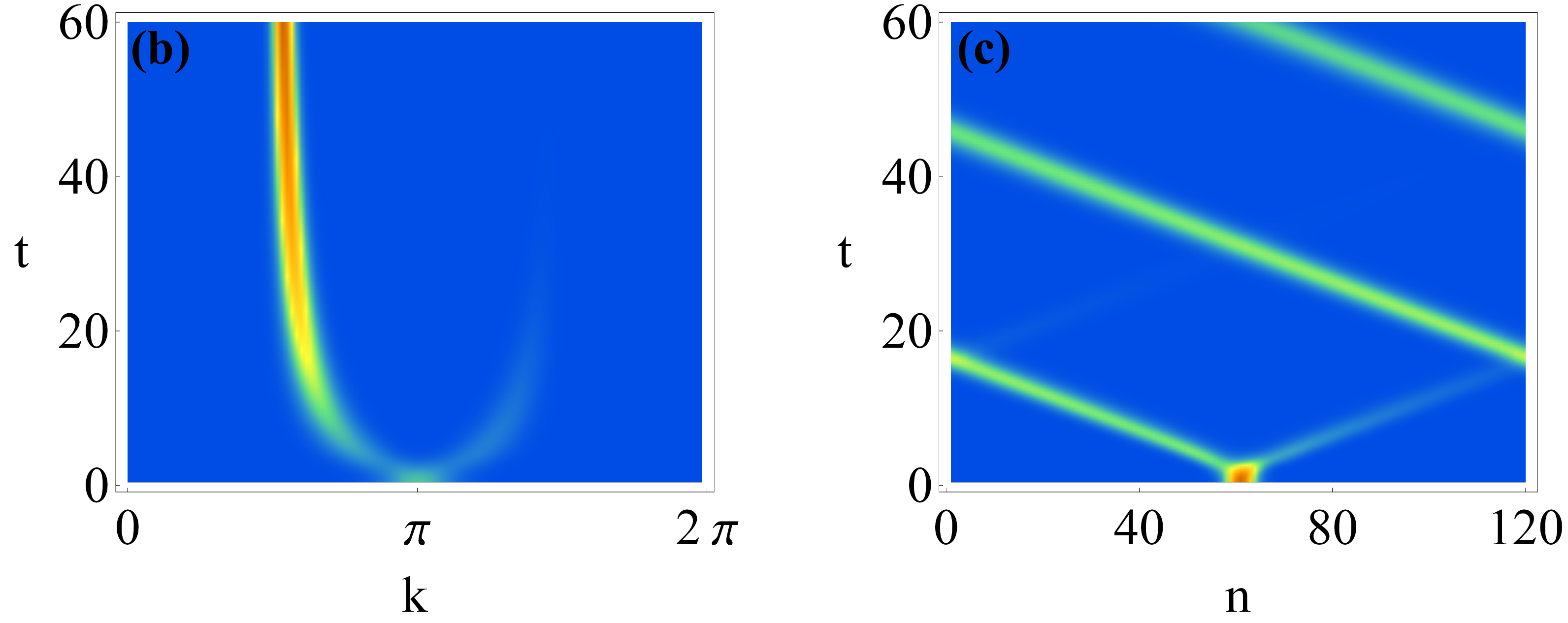}
    \caption{(a) In the model with explicit ATRS breaking [Eq. (\ref{supeq-sybre})], the originally coincident ATRS-protected loop spectra become separated, with $\lambda = 0.2$. (b)(c) Dynamical wavepacket evolutions in momentum space and real space under this ATRS-broken scenario. Other parameters are the same as those in the main text.}
    \label{supp-fig-symbre}
\end{figure}

\end{widetext}

\bibliography{ref.bib}
\end{document}